\documentclass[preprint,amsmath,amssymb,aps,pre,longbibliography]{revtex4-1}

\usepackage{graphicx}
\usepackage{dcolumn}
\usepackage{bm}
\usepackage{color}
\usepackage[utf8]{inputenc}

\newcommand{\dpar}[2]{\frac{\partial #1}{\partial #2}}

\begin{document}

\title{Application of the Widom insertion formula to transition rates in a lattice}

\author{M. A. Di Muro}
\affiliation{Instituto de Investigaciones F\'isicas de Mar del Plata (IFIMAR-CONICET), Departamento de F\'isica, Facultad de Ciencias Exactas y Naturales, Universidad Nacional de Mar del Plata, Funes 3350, 7600 Mar del Plata, Argentina}
\author{M. Hoyuelos}
\email{hoyuelos@mdp.edu.ar}
\affiliation{Instituto de Investigaciones F\'isicas de Mar del Plata (IFIMAR-CONICET), Departamento de F\'isica, Facultad de Ciencias Exactas y Naturales, Universidad Nacional de Mar del Plata, Funes 3350, 7600 Mar del Plata, Argentina}

\begin{abstract}
We consider diffusion of particles on a lattice in the so-called dynamical mean-field regime (memory effects are neglected). Interactions are local, that is, only among particles at the same lattice site. It is shown that a statistical mechanics analysis that combines detailed balance and Widom's insertion formula allows for the derivation of an expression for transition rates in terms of the excess chemical potential. The rates reproduce the known dependence of self-diffusivity as the inverse of the thermodynamic factor. Soft-core interactions and general forms of the excess chemical potential (linear, quadratic, and cubic with the density) are considered.
\end{abstract}


\maketitle

\section{Introduction}

The study of transport processes in a perfect lattice is of fundamental importance as a first step to understand, for example, diffusion in more complex and realistic systems, such as surfaces and solids \cite{ala,gomer,antczak,paul,mehrer}. One of the simplest models of this kind is the Langmuir gas \cite{kutner,kehr}, (see also \cite[Sec.\ 7.1]{hill} and \cite[Sec.\ 2.6.2.3]{gomer}) characterized by site exclusion due to hard core interaction; only one particle is allowed at each lattice site. Even in this case it is difficult to derive a closed analytical result for the tracer diffusivity \cite{ala,kutner,ferrando,hjelt}, the main difficulty being the presence of memory effects: backward jumps are more probable than jumps in other directions because when a particle moves it leaves behind an empty site.

We are interested in general interactions macroscopically represented by the excess chemical potential $\mu_{\text{ex}}$, a function of temperature $T$ and density $\rho$, with the limitation that there is not a phase transition. In the limit of small concentration, $\rho\rightarrow 0$, interactions can be neglected and $\mu_{\text{ex}}$ vanishes. For diffusion on surfaces and in solids, the Darken equation \cite{darken,ala,gomer,mehrer} gives a connection between the tracer and the collective diffusion coefficients, that is, between the diffusion of a tagged particle and the diffusion produced by a concentration gradient. The connection is given through the so-called thermodynamic factor, defined as $\Gamma = \beta \frac{\partial \mu}{\partial \ln \rho}$, where $\mu$ is the chemical potential and $\beta = (k_B T)^{-1}$; or, in terms of the excess chemical potential, as $\Gamma = 1 + \beta \rho \frac{\partial \mu_{\text{ex}}}{\partial \rho}$ (see, e.g., \cite[Sec.\ 2.6]{gomer}). The Darken equation manifests the decisive importance of the thermodynamic factor in the description of diffusion processes. It can be shown that the thermodynamic factor is directly related to particle number fluctuations.

A correction factor has to be included if memory effects are present. We consider the dynamical mean-field (DMF) regime \cite{ala} in which memory effects can be neglected. This approximation holds when there are many particles in each site (e.g., for soft-core interaction instead of hard core); in this case, the jump of one particle is a small perturbation of the initial state.

The Widom insertion formula \cite{widom} relates the excess chemical potential with the insertion energy, that is, the energy needed to insert one particle. The main purpose of this paper is to demonstrate that the Widom insertion formula, combined with detailed balance, provides relevant information for transition rates. Knowledge of transition rates is necessary when performing nonequilibrium simulations with kinetic Monte Carlo (if only the energy change is known, transition rates are generally obtained with Glauber or Metropolis algorithms, which do not guarantee a correct timescale for simulations out of equilibrium). Moreover, with transition rates, the DMF tracer diffusivity is immediately obtained. The results are checked with numerical simulations for soft-core interaction and for excess chemical potential linear, quadratic, and cubic with density.

The approach sketched in this Introduction is based on previous work on diffusion in solids  \cite{mdp,dipietro2}, that made possible to reproduce, using a formula with three free parameters, experimental results of the intrinsic diffusivity of different binary mixtures \cite{dipietro3}. Here, the intention is to reformulate and generalize calculations, partially present in those references, starting from fundamental concepts of statistical mechanics, and to numerically check some results. The systems analyzed are discrete: particles jump between neighboring sites in a $d$-dimensional lattice. Different local interactions, among particles in the same cell, are considered; interactions between particles in different cells are neglected.

The paper is organized as follows. In Sec.\ \ref{sec:theory}, the theory is developed. From detailed balance, transition rates of a tagged particle are written in terms of the configuration energy (Sec.\ \ref{sec:detbal}). The Widom insertion formula (Sec.\ \ref{sec:Widom}), is used to obtain an expression for transition rates, the result includes an undetermined function of the average concentration. From transition rates, collective diffusion and DMF tracer diffusion coefficients are obtained (Sec.\ \ref{sec:dif}). Some calculations are written in the Appendices in order to present a clearer picture of the main lines of reasoning. In Sec.\ \ref{sec:results}, numerical results are compared with the theory. Applications to surface diffusion are discussed in Sec.\ \ref{sec:surf}. Summary and conclusions are presented in Sec.\ \ref{sec:conclusions}.

\section{Theory}
\label{sec:theory}

We have a $d$-dimensional lattice. Each site, identified with index $i$, is a cell with $n_i$ particles. A generic cell is considered as an open system connected with a reservoir, constituted by the rest of the lattice, that imposes a temperature $T$ and chemical potential $\mu$. There are $\Omega$ microscopic states for one particle in a cell; $\Omega$ can be taken as a measure of the cell's volume, and we define the density as $n_i/\Omega$. Density spatial and temporal variations are smooth, hence local thermal equilibrium holds. The model can also be interpreted as the discretization of a continuous system where the cell size is much larger than the interaction range, and the interaction energy at cell walls is neglected with respect to the bulk.

\subsection{Detailed balance}
\label{sec:detbal}

Let us consider a jump process between cells 1 and 2, that have $n_1$ and $n_2$ particles, respectively. The initial state, $A$, is determined by the number of particles in the two cells: $A=\{n_1,n_2\}$. State $A$ undergoes a transition to state $B=\{n_1-1,n_2+1\}$, in which cells 1 and 2 have $n_1-1$ and $n_2+1$ particles. The transition rate from $A$ to $B$ is $W_{A,B}$, and $W_{B,A}$ is the corresponding rate for the inverse process.

The detailed balance relationship is
\begin{equation}\label{eq:detbalall}
P_A\, W_{A,B} = P_B\, W_{B,A},
\end{equation}
where $P_A$ and $P_B$ are the probabilities of states $A$ and $B$. Local thermal equilibrium is a sufficient condition for the validity of this relationship. 

The canonical partition function of $n$ particles in a cell is $\mathcal{Z}(n,T,\Omega)$, or $\mathcal{Z}_n$ for brevity. If the lattice is the discretization of a continuous system, then $\mathcal{Z}_n = \mathcal{Z}_{0,n} \, \langle e^{-\beta \mathcal{U}(\mathbf{q}_1,\dots,\mathbf{q}_n)} \rangle^0$, where $\mathcal{U}(\mathbf{q}_1,\dots,\mathbf{q}_n)$ is the interaction energy of $n$ particles at positions $\mathbf{q}_1,\dots,\mathbf{q}_n$ in the cell; see, e.g. (\cite[Sec.\ 5.1]{kardar}). The average $\langle \  \rangle^0$ is computed with the probability distribution of non-interacting particles; $\mathcal{Z}_{0,n}$ is the partition function of the ideal gas, given by $V^n/(\lambda^{3n} n!)$, where $\lambda$ is the thermal de Broglie wavelength and $V$ is the cell's volume. For particles in a lattice, the state is given by their positions, there is no velocity, and the canonical partition function is
\begin{equation}\label{eq:Z}
\mathcal{Z}_n = \sum_{\omega} e^{-\beta E_\omega} = \frac{\Omega^n}{n!} \langle e^{-\beta E_\omega}\rangle,
\end{equation} 
where the sum is over all microstates of $n$ particles, and $E_\omega$ is the interaction energy of microstate $\omega$. The sum is replaced by the total number of microstates, $\Omega^n/n!$, times the canonical average of the Boltzmann factor. Let us notice that, in the lattice, $\Omega$ plays the role of $V/\lambda^3$ for the continuous system. In the limit of small concentration, interactions are neglected and the canonical partition function is $\mathcal{Z}_{0,n} = \Omega^n/n!$.

The grand partition function of a cell is
\begin{equation}\label{eq:Q}
\mathcal{Q}(\mu,T,\Omega) = \sum_{n=0}^\infty e^{\beta \mu n} \mathcal{Z}_n.
\end{equation}
The probability $P_n$ of having $n$ particles in a cell is 
$P_n = e^{\beta \mu n} \mathcal{Z}_n/\mathcal{Q}$ and the probabilities of state $A$ and $B$ are $P_A = P_{n_1} P_{n_2}$ and $P_B = P_{n_1-1} P_{n_2+1}$ (this approximation is equivalent to writing the partition function of two cells with $n_1$ and $n_2$ particles as the product $\mathcal{Z}_{n_1} \mathcal{Z}_{n_2}$, since interaction energy at the walls is neglected). Then, detailed balance \eqref{eq:detbalall} implies
\begin{equation}\label{eq:detbala2}
\mathcal{Z}_{n_1}\mathcal{Z}_{n_2}\, W_{A,B} = \mathcal{Z}_{n_1-1}\mathcal{Z}_{n_2+1}\, W_{B,A}.
\end{equation}

It is useful to define the configuration energy of $n$ particles, $\phi_n$, as
\begin{equation}\label{eq:confen}
e^{-\beta \phi_n} = \langle e^{-\beta E_\omega} \rangle = \frac{\mathcal{Z}_n}{\mathcal{Z}_{0,n}}.
\end{equation}
In the thermodynamic limit we have that $\mathcal{Z}_n \stackrel{TL}{=} e^{-\beta F}$, with $F$ the free energy; symbol ``$\stackrel{TL}{=}$'' means that the equality holds in the thermodynamic limit. Therefore, $\phi \stackrel{TL}{=} F_\text{ex}$, with $F_\text{ex}$ the excess free energy. But it is important not to take the thermodynamic limit yet in order to keep nonextensive terms that turn out to be relevant for transition rates. Combining Eqs.\ \eqref{eq:Q} and \eqref{eq:confen}, the grand partition function can be written as
\begin{equation}\label{e.Q}
\mathcal{Q} = \sum_{n=0}^{\infty} \frac{\Omega^n}{n!} e^{-\beta\phi_n} e^{\beta\mu n}.
\end{equation}

Now, using the configuration energy, the detailed balance relationship \eqref{eq:detbala2} is
\begin{equation}\label{eq:ww}
\frac{W_{A,B}}{W_{B,A}} = \frac{e^{-\beta(\phi_{n_2+1}-\phi_{n_2})}}{e^{-\beta(\phi_{n_1}-\phi_{n_1-1})}} \frac{\mathcal{Z}_{0,n_1-1} \mathcal{Z}_{0,n_2+1}}{\mathcal{Z}_{0,n_1} \mathcal{Z}_{0,n_2}} = \frac{e^{-\beta(\phi_{n_2+1}-\phi_{n_2})}}{e^{-\beta(\phi_{n_1}-\phi_{n_1-1})}} \frac{n_1}{n_2+1}.
\end{equation}

The rate $W_{A,B}$ refers to the transition of one particle from cell 1 to cell 2; the jumping particle is any of those present in cell 1. For the description of tracer diffusion we need, instead, the transition rate of one tagged particle; the rate for one specific particle in cell 1 is $W_{A,B}/n_1$ since all particles are equivalent. Let us define $W_{n_1,n_2}$ as the transition rate for one tagged particle that jumps from cell 1 to cell 2, with $n_1$ and $n_2$ particles in each cell; the order of subscripts in $W_{n_1,n_2}$ indicates the direction of the jump. Then, $W_{n_1,n_2} = W_{A,B}/n_1$ and $W_{n_2+1,n_1-1} = W_{B,A}/(n_2+1)$, and Eq.\ \eqref{eq:ww} becomes
\begin{equation}\label{eq:wwt}
W_{n_1,n_2}\, e^{-\beta(\phi_{n_1}-\phi_{n_1-1})}
= W_{n_2+1,n_1-1}\, e^{-\beta(\phi_{n_2+1}-\phi_{n_2})}.
\end{equation}

\subsection{Widom insertion formula and transition rates}
\label{sec:Widom}

The Widom insertion formula  (\cite{widom}, see also \cite[p.\ 30]{hansen2}) is a relationship between the excess chemical potential, $\mu_{\text{ex}}$, and the interaction energy needed to insert one additional particle. It can be written as
\begin{equation}
e^{-\beta \mu_\text{ex}} = \langle e^{-\beta \,\Delta \phi_n} \rangle,
\label{eq:expexp}
\end{equation} 
where $\Delta \phi_n = \phi_{n+1}-\phi_n$ and the angular brackets represent the average in the grand canonical ensemble; see Appendix A for a derivation. Equation\ \eqref{eq:expexp} in the thermodynamic limit implies that $\phi' \stackrel{TL}{=} \mu_\text{ex}$, a result that, of course, is consistent with $\phi \stackrel{TL}{=} F_\text{ex}$. As usual in thermodynamics, $\phi_n$ is taken as a continuous function of $n$. The following notation is used to indicate derivatives with respect to the number of particles: $\phi' = \left.\frac{\partial \phi_n}{\partial n}\right|_{n=\bar{n}}$. Whenever $\phi$ or its derivatives are written without subindex, it is assumed that they are evaluated at the average number of particles, $\bar{n}$.

Using \eqref{eq:expexp}, it can be shown that (see Appendix B),
\begin{align}
\phi_{n_2+1}-\phi_{n_2} &= \mu_{\text{ex},n_2} + \varepsilon_{n_2} + \text{h.t.} \label{eq:Dphi2} \\
\phi_{n_1}-\phi_{n_1-1} &= \mu_{\text{ex},n_1} + \varepsilon_{n_1} + \text{h.t.} \label{eq:Dphi1}
\end{align}
with 
\begin{align}
\varepsilon_{n_2} &= - \frac{1}{2\beta} \frac{\Gamma'_{n_2}}{\Gamma_{n_2}} + \mu'_{\text{ex},n_2}/2 \label{eq:eps2} \\
\varepsilon_{n_1} &= - \frac{1}{2\beta} \frac{\Gamma'_{n_1}}{\Gamma_{n_1}} - \mu'_{\text{ex},n_1}/2, \label{eq:eps1}
\end{align}
where $\mu_{\text{ex},n_i}\sim \mathcal{O}(\Omega^0)$ and $\varepsilon_{n_i} \sim \mathcal{O}(\Omega^{-1})$ (each time a derivative with respect to $n_i$ is applied, the power order in $\Omega$ is reduced by 1). Higher order terms of $1/\Omega$ are represented by ``h.t.'' in \eqref{eq:Dphi2} and \eqref{eq:Dphi1}. Replacing \eqref{eq:Dphi2} and \eqref{eq:Dphi1} in \eqref{eq:wwt}, we obtain
\begin{equation}\label{eq:db0}
W_{n_1,n_2}\, e^{-\beta(\mu_{\text{ex},n_1} + \varepsilon_{n_1} + \text{h.t.})}
= W_{n_2+1,n_1-1}\, e^{-\beta(\mu_{\text{ex},n_2} + \varepsilon_{n_2} + \text{h.t.})}.
\end{equation}
Writing $W_{n_2+1,n_1-1} = W_{n_2,n_1} + \partial_{n_2} W_{n_2,n_1} - \partial_{n_1} W_{n_2,n_1} + \text{h.t.}$, we have
\begin{align}\label{eq:db1}
W_{n_1,n_2}\, & e^{-\beta\mu_{\text{ex},n_1}}(1 - \beta \varepsilon_{n_1} + \text{h.t.}) = \nonumber \\
& (\underbrace{W_{n_2,n_1}}_{\mathcal{O}(\Omega^0)} + \underbrace{\partial_{n_2} W_{n_2,n_1} - \partial_{n_1} W_{n_2,n_1} - \beta \varepsilon_{n_2} W_{n_2,n_1}}_{\mathcal{O}(\Omega^{-1})} + \text{h.t.})\, e^{-\beta\mu_{\text{ex},n_2}}.
\end{align}
Terms at different orders can be separated:
\begin{align}
\mathcal{O}(\Omega^0): \quad &W_{n_1,n_2} e^{-\beta\mu_{\text{ex},n_1}} = W_{n_2,n_1} e^{-\beta\mu_{\text{ex},n_2}} \label{eq:ov0} \\
\mathcal{O}(\Omega^{-1}): \quad & - \beta \varepsilon_{n_1} W_{n_1,n_2}e^{-\beta\mu_{\text{ex},n_1}} = \nonumber \\
& \qquad (\partial_{n_2} W_{n_2,n_1} - \partial_{n_1} W_{n_2,n_1} - \beta \varepsilon_{n_2} W_{n_2,n_1})\, e^{-\beta\mu_{\text{ex},n_2}}. \label{eq:ovm1}
\end{align}
Let us notice that the main idea in this procedure is to separate orders 0 and $1$, the orders at which we have information, but higher order terms are not neglected. Using \eqref{eq:ov0} in \eqref{eq:ovm1} we get
\begin{align}\label{eq:db2}
(\partial_{n_2} - \partial_{n_1}) \ln W_{n_2,n_1} &= \beta(\varepsilon_{n_2} - \varepsilon_{n_1}) \nonumber \\
&= -\frac{\Gamma_{n_2}'}{2\Gamma_{n_2}} + \frac{\beta}{2}\mu_{\text{ex},n_2}' + \frac{\Gamma_{n_1}'}{2\Gamma_{n_1}} + \frac{\beta}{2}\mu_{\text{ex},n_1}',
\end{align}
where expressions \eqref{eq:eps2} and \eqref{eq:eps1} for $\varepsilon_{n_i}$ were used in the last line.

Let us define $\nu_{n_2,n_1}$ such that
\begin{equation}\label{eq:defw}
\ln W_{n_2,n_1}-\ln \nu_{n_2,n_1} = - \frac{1}{2} \ln \Gamma_{n_2} + \frac{\beta}{2} \mu_{\text{ex},n_2} - \frac{1}{2} \ln \Gamma_{n_1} - \frac{\beta}{2} \mu_{\text{ex},n_1},
\end{equation}
where the right-hand side is defined in such a way that, when operator $\partial_{n_2}-\partial_{n_1}$ is applied, the right-hand side of Eq.\ \eqref{eq:db2} is obtained. Equivalently,
\begin{equation}\label{eq:defw2}
W_{n_2,n_1} = \nu_{n_2,n_1} \frac{1}{(\Gamma_{n_2}\Gamma_{n_1})^{1/2}} \frac{e^{\beta \mu_{\text{ex},n_2}/2}}{e^{\beta \mu_{\text{ex},n_1}/2}}.
\end{equation}
The expression for $W_{n_1,n_2}$ is obtained by exchanging $n_1 \leftrightarrow n_2$. Using this ansatz in Eqs.\ \eqref{eq:ov0} and \eqref{eq:db2} we obtain the following conditions for $\nu_{n_2,n_1}$:
\begin{align}
\nu_{n_2,n_1} &= \nu_{n_1,n_2} \\
\partial_{n_2}\ln \nu_{n_2,n_1} &= \partial_{n_1}\ln \nu_{n_2,n_1}.
\end{align}
The solution of these equations is a function that depends on the sum $n_1 + n_2$. Then, we can write $\nu_{n_2,n_1} = \nu_{n_2+n_1}$ and the transition rate is
\begin{equation}\label{eq:defw3}
W_{n_2,n_1} = \nu_{n_2+n_1} \underbrace{\frac{e^{-\beta \mu_{\text{ex},n_1}/2}}{\Gamma_{n_1}^{1/2}}}_{\psi_{n_1}} \underbrace{\frac{e^{\beta \mu_{\text{ex},n_2}/2}}{\Gamma_{n_2}^{1/2}}}_{\varphi_{n_2}}.
\end{equation}
We arrived at an expression for the transition rate that is the product of $\nu_{n_1+n_2}$ times two functions, $\psi_{n_1}$ and $\varphi_{n_2}$, that depend on $n_1$ and $n_2$ respectively. Both, $\psi_{n_1}$ and $\varphi_{n_2}$, tend to 1 in the limit of small concentration. This is the form of the transition rate that can be deduced taking advantage of the information provided by the Widom insertion formula. Function $\nu_{n_1+n_2}$ is unknown, but now we can advance with a physical interpretation. Since $\nu$ depends on $n_1+n_2$, it corresponds to an effect of the average concentration of both cells. The average concentration modifies, for example, the substratum for diffusion on a surface or in a solid, that is, it modifies the activation energy (the energy that a particle has to overcome to start a jump \cite[Sec.\ 5.3.5]{paul}). This type of information depends on microscopic specific characteristics of the system and, as expected, cannot be deduced with the coarse-grained general approach that is carried out here. In the examples used for numerical simulations, in Sec.\ \ref{sec:results}, a constant value of $\nu$ is assumed. For surface diffusion, $\nu$ depends, in general, on concentration (see Sec.\ \ref{sec:surf}). Nevertheless, the fact that $\nu$ depends on the sum $n_1+n_2$ implies that it can not be a function of the excess chemical potential, because $\mu_{\text{ex}}$ depends either on $n_1$ or $n_2$, not on the sum, since particles in different cells do not interact, and $\mu_{\text{ex}}$ is not extensive.

\subsection{Diffusivity}
\label{sec:dif}

Diffusion processes are mainly characterized by two coefficients. The collective diffusion coefficient, $D_c$, gives the decay rate of long wavelength fluctuations of particle concentration. More specifically, it is the coefficient that relates particle current, $J$, with concentration gradient in the first Fick's law. On the other hand, the single-particle or tracer diffusion coefficient is defined in terms of the mean-square displacement of one tagged particle, $D = \langle \Delta x^2\rangle/2t$, for large values of time $t$; for simplicity we consider diffusion in one direction, along the $x$ axis. Both coefficients are, in general, different; they coincide when interactions can be neglected. If the tagged particle interacts with particles of the same type, the tracer diffusivity is equivalent to the self-diffusion coefficient.

Let us consider the current in one direction and a smooth spatial variation of the linear concentration $c_i = n_i/a$, where $a$ is the cell's size. The particle current between a pair of generic cells 1 and 2 is
\begin{equation}\label{eq:J}
J = n_1 W_{n_1,n_2} - n_2 W_{n_2,n_1},
\end{equation}
where $n_1$ and $n_2$ are similar to $\bar{n}$. Using \eqref{eq:defw3},
\begin{align}
J &= \frac{\nu_{n_1+n_2}}{(\Gamma_{n_1}\Gamma_{n_2})^{1/2}}(n_1 e^{-\beta \Delta \mu_{\text{ex}}/2} - n_2 e^{\beta \Delta \mu_{\text{ex}}/2}) \nonumber \\
&\simeq \frac{\nu}{\Gamma} [n_1 - n_2 - \beta (n_1+n_2)\Delta \mu_{\text{ex}}/2] \nonumber \\
&\simeq - \frac{\nu}{\Gamma}\Delta n \left[1 + \beta \bar{n} \frac{\Delta\mu_\text{ex}}{\Delta n}\right] = -\nu\, \Delta n \nonumber\\
&= - \nu a^2 \frac{\Delta c}{a},
\label{eq:J2}
\end{align}
where $\Delta n = n_2-n_1$ and $\Delta \mu_{\text{ex}} = \mu_{\text{ex},n_2}-\mu_{\text{ex},n_1}$.
The proportionality factor between current and concentration gradient is, as mentioned before, the collective diffusion coefficient, then
\begin{equation}\label{eq:Dc}
D_c = \nu a^2.
\end{equation}

Let us consider the DMF approximation for tracer diffusivity. We denote the tracer diffusivity by $D^\text{MF}$ to indicate that memory effects are neglected. The tagged particle performs a random walk with an average jump rate $W$ and jump size $a$. The diffusion coefficient is obtained in the continuous limit of the random walk, and the result is $D^\text{MF} = W a^2$ (see \cite[Sec.\ 3.8.2]{gardiner}). From Eq.\ \eqref{eq:defw3}, the average jump rate in equilibrium is $W= \nu/\Gamma$; then
\begin{equation}\label{eq:Dt}
D^\text{MF} = \nu a^2/ \Gamma.
\end{equation} 
Combining Eqs.\ \eqref{eq:Dc} and \eqref{eq:Dt}, we recover the Darken equation \cite{darken}:
\begin{equation}\label{eq:darken}
D^\text{MF} = D_c/ \Gamma,
\end{equation}
that is known to hold when memory effects are neglected \cite{ala}. The present procedure provides additional information, since we have the transition rates and two separate expressions for collective and tracer diffusivities. 

Now, we can interpret the meaning of all terms in the transition rate \eqref{eq:defw3}: $\nu_{n_1+n_2}$ is the effect of the substratum; $(\Gamma_{n_1}\Gamma_{n_2})^{-1/2}$ gives the dependence on the thermodynamic factor that appears in the Darken equation; and $e^{-\beta\Delta \mu_{\text{ex}}/2}$ is a Boltzmann factor with $\mu_{\text{ex}}$ corresponding to a mean field potential for one tagged particle. Citing \cite[p.\ 29]{taglia}: ``the excess chemical potential can be thought of as an effective mean-field potential acting on the particle due to the presence of other particles and external forces.''

\section{Comparison with numerical results}
\label{sec:results}

\subsection{Soft core}
\label{sec:soft-core}

Soft core is an illustrative example. Partition function, configuration energy, and excess chemical potential can be obtained (in other examples we assume that only the excess chemical potential is known). We define soft core as a generalization of hard core: instead of only one particle per lattice site, the maximum number of particles is an arbitrary number $\Omega$. The grand partition function for hard core is that of the Fermi-Dirac distribution: $\mathcal{Q}_\text{HC} = 1 + e^{\beta\mu}$. The soft-core partition function is
\begin{equation}\label{eq:softpartition}
\mathcal{Q} = (1 + e^{\beta\mu})^\Omega = \sum_{n=0}^{\Omega} \frac{\Omega!}{(\Omega-n)! n!} e^{\beta\mu n}.
\end{equation}
Hard core is recovered when $\Omega=1$. With this definition, if $\bar{n}_\text{HC}$ is the mean number of particles for hard core, the mean number of particles for any $\Omega$ is $\bar{n}=\Omega\,\bar{n}_\text{HC}$. Comparing with \eqref{e.Q},
\begin{equation}\label{eq:phisoft}
e^{-\beta\phi_n} = \frac{\Omega!}{\Omega^n (\Omega-n)!}
\end{equation}
and
\begin{equation}\label{eq:softDphi}
e^{-\beta(\phi_{n+1}-\phi_n)} = 1 - \frac{n}{\Omega}.
\end{equation}
Using the Widom insertion formula \eqref{eq:expexp}, the excess chemical potential is
\begin{equation}\label{eq:softex}
e^{-\beta\mu_{\text{ex}}} = 1 - \rho,
\end{equation}
with $\rho=\bar{n}/\Omega$, and the thermodynamic factor is
\begin{equation}\label{eq:softtf}
\Gamma = \frac{1}{1-\rho}
\end{equation}
[see Eq.\ (2.106) in Ref.\ \cite{gomer}]. With these expressions evaluated at $n_1$ and $n_2$, we obtain that the transition rate \eqref{eq:defw3} depends only on the number of particles in the destination cell:
\begin{equation}\label{eq:softW}
W_{n_1,n_2} = \nu (1-\rho_2),
\end{equation}
with $\rho_2 = n_2/\Omega$. That is, the transition probability to a site is proportional to the available space, given by $1-\rho_2$. The DMF tracer diffusivity \eqref{eq:Dt} is
\begin{equation}\label{eq:softDt}
D^\text{MF}/\nu a^2 = 1 - \rho.
\end{equation}
This result is numerically reproduced in Fig.\ \ref{fig:soft} for two dimensions and for $\Omega=100$. In the same figure, the inset shows the collective diffusion coefficient $D_c$ against concentration to verify Eq.\ \eqref{eq:Dc}. These are well-known results that are reproduced here in order to verify the validity of the procedure. The collective diffusivity is numerically calculated in the following way. The system has size $L_x\times L_y$; periodic boundary conditions are used in the $y$ direction; a constant flux $J$ of incoming particles is applied at $x=0$ and for all $y$; at $x=L_x$ particles are removed. The system evolves until the stationary state is reached. At this state, the density has a decreasing gradient in the $x$ direction, $\frac{\partial \rho}{\partial x}$, that, in the present case, is independent of position (or density). The collective diffusivity is obtained from $D_c = -J/\frac{\partial\rho}{\partial x}$. In the simulations, $L_x$ should be large enough to have small values of $\beta \Delta \mu_{\text{ex}}$ in the stationary state and in the whole system. The same procedure is used for the cases analyzed in the next sections.

\subsection{Effective boson interaction}

\begin{figure}[p]
  \includegraphics[width=.85\columnwidth]{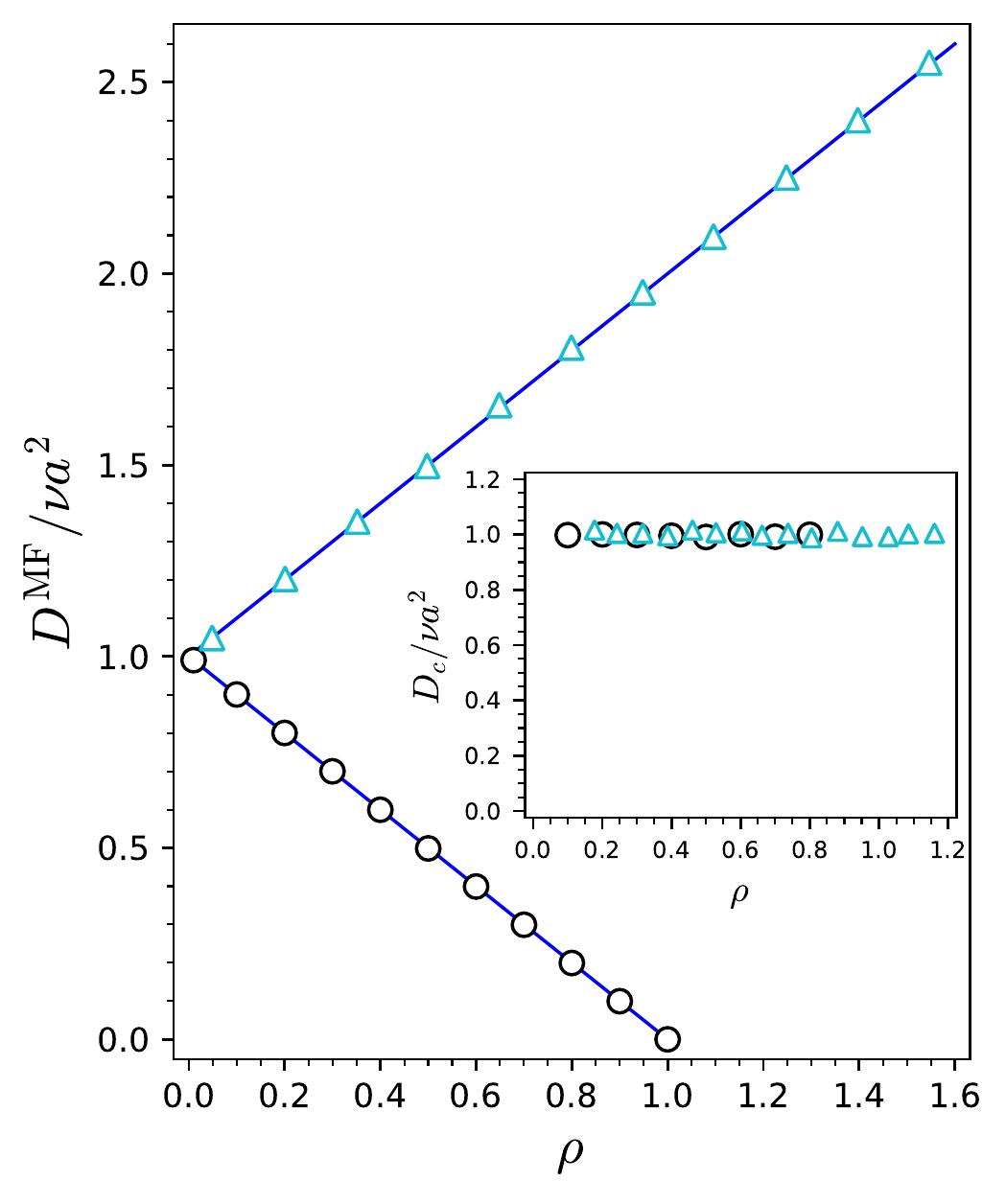}
	\caption{Numerical results of the normalized DMF tracer diffusivity against density for soft-core interaction (circles) and effective boson interaction (triangles), with $\Omega=100$; lines correspond to Eqs.\ \eqref{eq:softDt} and \eqref{e.tracerbe}, respectively. Parameters of the Monte Carlo simulation for soft core:  between $300$ and $5000$ realizations were performed depending on the density value, each consisting of $1000$ Monte Carlo time steps, in a $100\times100$ square lattice. The inset shows the normalized collective diffusivity against density (in a $1000\times100$ square lattice for soft core); it has a constant value as predicted by Eq.\ \eqref{eq:Dc} for both interactions. Numerical data for effective boson interaction were taken from Ref.\ \cite{suarez}.}
	\label{fig:soft}
\end{figure}

A system of classical particles that reproduce Bose-Einstein statistics is considered in this section. Since there are $\Omega$ microstates for one particle, the grand partition function is
\begin{equation}\label{e.QBE}
\mathcal{Q} = \left( \frac{1}{1 - e^{\beta \mu}} \right)^\Omega = \sum_{n=0}^\infty \frac{\Omega (\Omega+1)\cdots (\Omega+n-1)}{n!} e^{\beta\mu n}
\end{equation}
where the binomial series was used. Comparing with \eqref{e.Q} we get
\begin{equation}\label{e.confbe}
e^{-\beta \phi_n} = \frac{\Omega (\Omega+1)\cdots (\Omega+n-1)}{\Omega^n},
\end{equation}
and 
\begin{equation}\label{e.difbe}
e^{-\beta (\phi_{n+1}-\phi_n)} = 1 + \frac{n}{\Omega}. 
\end{equation}
Then, using Eq.\ \eqref{eq:expexp}, the excess chemical potential is
\begin{equation}\label{e.muexbe}
\mu_{\text{ex}} = -\beta^{-1} \ln(1+\rho).
\end{equation}
This example is qualitatively different from soft core, since the effective interaction that reproduces boson's statistics in a classical system is attractive, resulting in an excess chemical potential that decreases with concentration. The corresponding thermodynamic factor is
\begin{equation}\label{e.gammabe}
\Gamma = \frac{1}{1+\rho},
\end{equation}
and the transition rate is
\begin{equation}\label{e.Wbe}
W_{n_1,n_2} = \nu (1 + \rho_2).
\end{equation}
Including the effect of an external force in the transition rate, a closed system in equilibrium has Bose-Einstein statistics (see \cite{suarez,dipietro4}). As for soft core, the transition rate depends only on concentration in the destination cell. The DMF tracer diffusivity is
\begin{equation}\label{e.tracerbe}
D^\text{MF}/\nu a^2 = 1 + \rho.
\end{equation}
Numerical results shown in Fig.\ \ref{fig:soft} verify this equation for $D^\text{MF}$, and also Eq.\ \eqref{eq:Dc} for $D_c$.

\subsection{Linear, quadratic and cubic excess chemical potential}

In order to calculate transition rates \eqref{eq:defw3}, only the excess chemical potential is needed. In this section we consider
\begin{equation}\label{eq:chempot}
\beta\mu_{\text{ex}} = \rho^k
\end{equation}
with $k=1$, 2, and 3, so that interactions, and the excess chemical potential, become relevant when the density, $\rho=\bar{n}/\Omega$, is of order 1 or larger. In order to avoid memory effects, a value $\Omega=100$ was used in the simulations; in this way, when the number of particles is of order 100, one jump represents a small perturbation and the DMF regime holds.

\begin{figure}[h]
	\includegraphics[width=0.97\columnwidth]{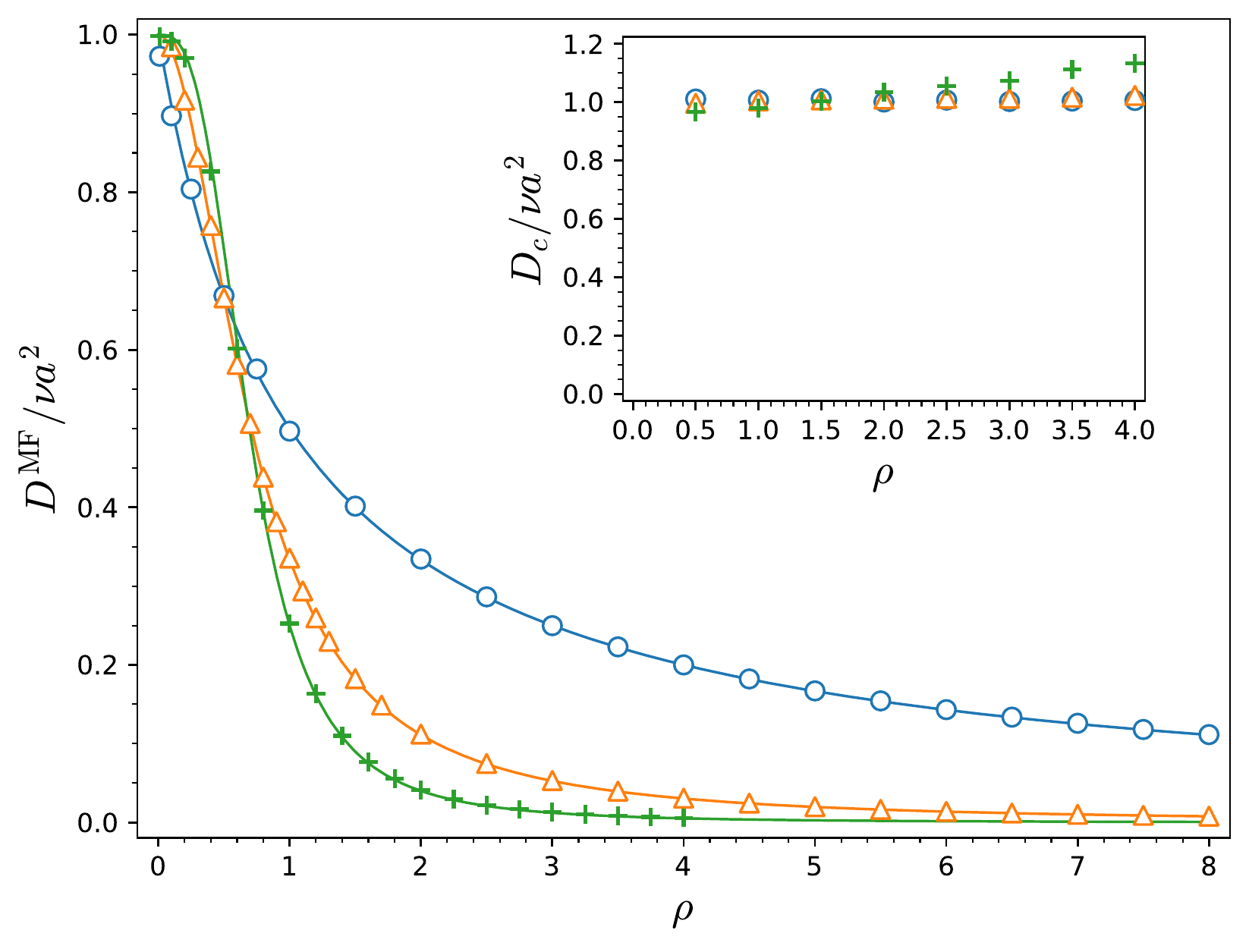}
	\caption{Numerical results of the normalized DMF tracer diffusivity against density for $\beta\mu_{\text{ex}}=\rho^k$ with $k=1$ (blue circles), 2 (orange triangles), and 3 (green plus sign) in a $50\times50$ square lattice, with $\Omega=100$. Curves correspond to Eq.\ \eqref{eq:trdiff}. Parameters of the Monte Carlo simulation: between $1000$ and $100$ realizations were performed depending on the density value, each consisting of $200$ Monte Carlo time steps, $a=1$ and $\nu=1/4$. Normalized collective diffusivity against density is shown in the inset for the three mentioned cases, in which the lattice size is $1000\times50$ for $k=1$ and $k=2$, and $10000\times5$ for $k=3$.  The result of an approximately constant value of $D_c$ verifies Eq.\ \eqref{eq:Dc}; there is a small deviation for $k=3$ originated in numerical difficulties to satisfy the condition of a small excess chemical potential variation between neighboring sites (it requires a much larger system length $L_x$ than in the other cases).}
	\label{fig:2}
\end{figure}

The thermodynamic factor is
\begin{equation}\label{eq:thermf}
\Gamma = 1+k\rho^k
\end{equation} 
and the transition rate is
\begin{equation}\label{eq:tr}
W_{n_1,n_2} = \nu \frac{e^{(\rho_1^k - \rho_2^k)/2}}{(1+k\rho_1^k)^{1/2} (1+k\rho_2^k)^{1/2}}.
\end{equation}
The resulting DMF tracer diffusivity is
\begin{equation}\label{eq:trdiff}
D^\text{MF}/\nu a^2 = \frac{1}{1+k\rho^k}.
\end{equation}
Figure\ \ref{fig:2} shows numerical results of $D^\text{MF}/\nu a^2$ against density for $k=1$, 2, and 3 in a two-dimensional lattice (details of the simulation in the figure caption). A good agreement with Eq.\ \eqref{eq:trdiff} is obtained. The inset contains numerical results of the normalized collective diffusion coefficient $D_c/\nu a^2$ as a function of density for the same cases, showing an approximately constant value equal to 1 in agreement with Eq.\ \eqref{eq:Dc}.

\section{Surface diffusion}
\label{sec:surf}

Applications of the results to diffusion on surfaces are discussed here. The DMF approximation for tracer diffusivity is not valid in general for this case. The so-called correlation factor, $f_t$, has to be included in order to take memory effects into account:
\begin{equation}\label{eq:Dft}
D = D^\text{MF} f_t.
\end{equation}
There is not a general method to obtain $f_t$; different approaches are described in, for example, \cite[Chap.\ 5]{paul} or \cite[Chap.\ 7]{mehrer} for diffusion in solids.
\newpage
The following expressions for tracer and collective diffusivity ($D$ and $D_c$) can be found in the literature on surface diffusion \cite{ala,gomer,reed}:
\begin{align}
D &= a^2 W f_t \label{e.Dlit} \\
D_c &= a^2 W \Gamma, \label{e.Dclit}
\end{align}
where $W$, the average jump rate, is a function of the coverage $\rho$. It is well known that $W\propto 1-\rho$ and $\Gamma=1/(1-\rho)$ for hard core interaction (Langmuir gas) \cite[Sec.\ 2.6.2.3]{gomer}, so that the dependence on $\rho$, or $\Gamma$, is canceled in the expression of $D_c$ for this case. As far as we know, there is not a general relationship between $W$ and $\Gamma$, for any interaction, in the literature on surface diffusion. We have shown that this relationship is $W=\nu/\Gamma$, where $\nu$ is, in general, a function of the concentration that cannot be written in terms of the excess chemical potential (or the thermodynamic factor). Equations\ \eqref{e.Dlit} and \eqref{e.Dclit} become
\begin{align}
D &= a^2 \nu f_t/\Gamma, \nonumber \\ 
D_c &= a^2 \nu. \nonumber 
\end{align}
Now we can interpret $\nu$ as the jump rate associated to the collective diffusion coefficient. These expressions are consistent with the results of Sec.\ \ref{sec:dif} [see Eqs.\ \eqref{eq:Dc} and \eqref{eq:Dt}]. The new information introduced was $W=\nu/\Gamma$.

In the examples of the previous section, a constant value of $\nu$ was assumed for the numerical test. As mentioned before, $\nu$ is not constant in general. It depends on the energy landscape that a particle has to overcome in order to jump between cells, and on geometric aspects such as the spatial distribution of energy wells of different depth. Also, energy barriers may depend on concentration. Therefore, even if $\nu$ is independent of $\mu_\text{ex}$, the collective diffusion coefficient may depend on concentration due to features of the energy landscape. A constant value of $\nu$ is an approximation useful to develop simplified models, but it fails in general for real systems. The result obtained here for the collective diffusion coefficient indicates that, if $D_c$ depends on concentration, this dependence is not a direct effect of the excess chemical potential but, instead, it is produced by microscopic details such as the modification of the substratum due to the presence of other particles or geometrical aspects of the energy landscape. This is a useful guide for seeking theoretical explanations for concentration-dependent collective diffusivity in more complex scenarios. An example is a variational method introduced in \cite{gortel} (see also \cite{badowski,minkowski2,minkowski}, where the method has been applied to the calculation of the collective diffusion coefficient of adsorbates in different surfaces).

\section{Summary and conclusions}
\label{sec:conclusions}

In summary, combining detailed balance and the Widom insertion formula, an equation for transition rates is obtained. Terms of different order in $\Omega$ (the number of microstates for one particle) are present in the detailed balance relationship. The procedure is based on separation of terms $\mathcal{O}(\Omega^0)$ and $\mathcal{O}(\Omega^{-1})$. The equation for the transition rate between two adjacent cells, with $n_1$ and $n_2$ particles, is proportional to three factors: an undetermined function representing substratum effects ($\nu_{n_1+n_2}$), the inverse of the thermodynamic factor [more precisely, $1/(\Gamma_{n_1}\Gamma_{n_2})^{1/2}$] that anticipates Darken equation, and a Boltzmann factor with the excess chemical potential ($e^{-\beta\Delta\mu_{\text{ex}}}$). A limitation of the theory is that the final result holds as long as there is not a phase transition, since $\Gamma$ vanishes in that case, and the expansion in terms of the particle number fluctuations used in Appendix B is no longer valid. The results were obtained assuming local interactions, only among particles in the same cell.

The present approach is intended to understand interaction effects on diffusion at a thermodynamic or macroscopic level, where interactions are represented by the excess chemical potential. Using the transition rates, we have shown that $\mu_{\text{ex}}$ has no effect on the collective diffusion coefficient, $D_c$, while the DMF tracer diffusivity, $D^\text{MF}$, is inversely proportional to the thermodynamic factor. Numerical simulations confirm that, for different functions of $\mu_{\text{ex}}$ against concentration, $D^\text{MF}$ changes but $D_c$ remains constant. Since the simulations were designed to check the effects of $\mu_{\text{ex}}$, parameter $\nu$ was assumed constant. Parameter $\nu$ represents microscopic details, it cannot be determined at a macroscopic description level in terms of $\mu_{\text{ex}}$, and, in general, depends on concentration. It is necessary to include microscopic details of the energy landscape to calculate $\nu$ or the collective diffusivity.
	
One important conclusion is that the Widom insertion formula provides relevant information for the determination of transition rates. Transition rates are required to perform kinetic Monte Carlo simulations of nonequilibrium regimes with the correct time scale.

\section*{Acknowledgments}
M.H. acknowledges discussions with H. M\'artin and M. Di Pietro Martínez that were useful for the development of these ideas. This work was partially supported by Consejo Nacional de Investigaciones Cient\'ificas y T\'ecnicas (CONICET, Argentina, Grant No. PIP 112 201501 00021 CO).

\section*{Appendix A}

A derivation of the Widom insertion formula in the grand canonical ensemble, Eq.\ \eqref{eq:expexp}, is presented in this appendix.

Using Eq.\ \eqref{e.Q}, the grand partition function is
\begin{equation}
\mathcal{Q} = \sum_{n=0}^\infty \frac{1}{n!} e^{\beta \mu n} \, e^{-\beta(\phi_n +\mu^\circ n)},
\label{granZ}
\end{equation}
with $\mu^\circ = -k_B T \ln(\Omega)$. First, let us notice that $\mathcal{Q}$ reproduces the behavior of the ideal system when interactions are neglected ($\phi_n=0$). In this case, from $\mathcal{Q}$ we obtain the following result for the mean number of particles:
\begin{equation}
\bar{n} = e^{\beta (\mu-\mu^\circ)} \qquad \text{(ideal case)}
\end{equation}
or $\mu = \mu^\circ+\beta^{-1}\ln \bar{n}$, i.e., the expression for the ideal chemical potential.

In the general case we have to include the excess chemical potential,
\begin{equation}
\bar{n} = e^{\beta (\mu-\mu^\circ)}\, e^{-\beta \mu_\text{ex}},
\label{eq:barnmu}
\end{equation}
and, from the grand partition function,
\begin{align}
\bar{n} &= 
 \frac{1}{\mathcal{Q}} \sum_{n=0}^{\infty} \frac{n}{n!} e^{-\beta(\phi_n + \mu^\circ n - \mu n)} \nonumber \\
&= \frac{e^{\beta (\mu-\mu^\circ)}}{\mathcal{Q}} \sum_{n=1}^{\infty} \frac{1}{(n-1)!} e^{-\beta[\phi_n + \mu^\circ (n-1) -\mu (n-1)]} \nonumber \\
&=  \frac{e^{\beta (\mu-\mu^\circ)}}{\mathcal{Q}} \sum_{m=0}^{\infty} \frac{1}{m!} e^{-\beta(\phi_{m+1} + \mu^\circ m - \mu m)} \nonumber \\
&=  \frac{e^{\beta (\mu-\mu^\circ)}}{\mathcal{Q}} \sum_{m=0}^{\infty} \frac{1}{m!} e^{-\beta(\phi_{m+1}-\phi_m)} e^{-\beta(\phi_m + \mu^\circ m - \mu m)} \nonumber \\
&= e^{\beta (\mu-\mu^\circ)} \langle e^{-\beta(\phi_{n+1}-\phi_n)} \rangle,
\label{eq:nmed1}
\end{align}
where the summation index was changed in the third line: $m=n-1$. Then, from \eqref{eq:barnmu} and \eqref{eq:nmed1} we have the Widom insertion formula
\begin{equation}
e^{-\beta \mu_\text{ex}} = \langle e^{-\beta \,\Delta \phi_n} \rangle, \nonumber
\end{equation} 
with $\Delta \phi_n = \phi_{n+1}-\phi_n$. Let us notice that the present derivation relies on the grand canonical ensemble average, while the canonical ensemble average is frequently used in the literature \cite[p.\ 30]{hansen2}.

\section*{Appendix B}

Expressions for $\phi_{n_2+1}-\phi_{n_2}$ and $\phi_{n_1}-\phi_{n_1-1}$ are derived in this appendix. The starting point is Eq.\ \eqref{eq:expexp}, $e^{-\beta \mu_\text{ex}} = \langle e^{-\beta \,\Delta \phi_n} \rangle$. We need an approximation for the average in the right hand side. 

We know that $\phi \stackrel{TL}{=} F_\text{ex}$ and $\phi' \stackrel{TL}{=} \mu_\text{ex}$. The purpose is to evaluate the difference $\phi'- \mu_\text{ex}$ up to order $\Omega^{-1}$. As mentioned before, $\Omega$ is a measure of the cell's volume.

Let us call $f(n)=e^{-\beta \,\Delta \phi_n}$. The number of particles $n$ is a stochastic variable with mean value $\bar{n}$ of order $\Omega$. We approximate
\begin{equation}\label{eq:fexp}
\langle f(n)\rangle = f(\bar{n}) + \frac{f''(\bar{n})}{2} \langle \Delta n^2\rangle + \text{h.t.},
\end{equation}
where $\Delta n = n-\bar{n}$, $\langle \Delta n\rangle =0$ and h.t.\ represents terms $\mathcal{O}(\Omega^{-2})$ or smaller. This expansion holds as long as there is no phase transition, since in that case the average squared fluctuations of particle number diverges. There is a relationship between fluctuations and thermodynamic factor, defined as $\Gamma = \beta\bar{n} \frac{\partial\mu}{\partial\bar{n}}=1+\beta \bar{n} \mu_\text{ex}'$; it is given by 
\begin{equation}\label{eq:dngamma}
\langle \Delta n^2 \rangle = \frac{1}{\beta^2} \frac{\partial^2 \ln \mathcal{Q}}{\partial\mu^2} = \frac{1}{\beta} \dpar{\bar{n}}{\mu}=\bar{n}/\Gamma.
\end{equation}
Using that $\Delta \phi_n = \phi_n' + \phi_n''/2 + \cdots$ (this expansion is obtained from the Taylor series of $\phi_{n+1}$ around $n$ with $\Delta n = 1$), and that $\phi_n\sim \mathcal{O}(\Omega)$, $\phi_n'\sim \mathcal{O}(\Omega^0)$, $\phi_n''\sim \mathcal{O}(\Omega^{-1})$, etc.,  we have
\begin{align}
f(\bar{n}) &=  e^{-\beta \phi'}(1-\beta \phi''/2) + \mathcal{O}(\Omega^{-2}) \\
f''(\bar{n}) &= e^{-\beta \phi'} \beta(\beta \phi''^2 - \phi''') + \mathcal{O}(\Omega^{-3})
\end{align}
Going back to the Widom insertion formula, $e^{-\beta \mu_\text{ex}} = \langle f(n) \rangle$, we have
\begin{equation}
e^{-\beta\mu_\text{ex}} = e^{-\beta \phi'}(1 + \beta \epsilon + \text{h.t.}),
\label{eq:Vbeps}
\end{equation}
with
\begin{equation}
\epsilon = -\frac{1}{2}\phi'' + \frac{1}{2}(\beta\phi''^2 - \phi''') \frac{\bar{n}}{\Gamma}.
\label{eq:epsphi}
\end{equation}
It can be seen that $\epsilon$ is of order $\Omega^{-1}$. Taking the logarithm of \eqref{eq:Vbeps} we have,
\begin{equation}\label{eq:phip}
\phi' = \mu_{\text{ex}} + \epsilon + \text{h.t.}
\end{equation}
The second and third derivatives of $\phi$ in \eqref{eq:epsphi} can be obtained from \eqref{eq:phip}: $\phi''=\mu_\text{ex}' + \mathcal{O}(\Omega^{-2})$ and $\phi'''=\mu_\text{ex}'' + \mathcal{O}(\Omega^{-3})$. Keeping the order $\Omega^{-1}$ in $\epsilon$, Eq.\ \eqref{eq:epsphi} is
\begin{align}
\epsilon &= -\frac{1}{2}\mu_\text{ex}' + \frac{1}{2}(\beta \mu_\text{ex}'^2 - \mu_\text{ex}'') \frac{\bar{n}}{\Gamma} \nonumber \\
&= -\frac{\mu_\text{ex}'+\bar{n}\mu_\text{ex}''}{2(1 + \beta\bar{n}\mu_\text{ex}')} = -\frac{1}{2\beta} \frac{\partial}{\partial \bar{n}} \ln(1 + \beta\bar{n}\mu_\text{ex}') \nonumber \\
&= -\frac{1}{2\beta} \frac{\partial}{\partial \bar{n}} \ln \Gamma. 
\label{eq:epsilon2}
\end{align}
Then,
\begin{equation}\label{eq:phinm}
\phi' = \mu_\text{ex} -\frac{1}{2\beta} \frac{\Gamma'}{\Gamma} + \text{h.t.} 
\end{equation}
The expression for a specific value of $n$ (instead of $\bar{n}$) should have the same form:
\begin{equation}
\phi'_n  = \mu_{\text{ex},n} - \frac{1}{2\beta} \frac{\Gamma'_n}{\Gamma_n} + \text{h.t.} 
\end{equation}
so that, when average is applied, Eq.\ \eqref{eq:phinm} is recovered (higher order terms, h.t., are different in both equations). Then, the second derivative of the configuration energy (that is used below) is $\phi''_n  = \mu_{\text{ex},n}' + \text{h.t.}$


We are interested in the differences $\phi_{n_2+1}-\phi_{n_2}$ and $\phi_{n_1}-\phi_{n_1-1}$ that appear in \eqref{eq:wwt}; they are
\begin{align}
\phi_{n_2+1}-\phi_{n_2} &= \phi_{n_2}' + \phi_{n_2}''/2 +\text{h.t.} = \mu_{\text{ex},n_2} - \frac{1}{2\beta} \frac{\Gamma'_{n_2}}{\Gamma_{n_2}} + \mu'_{\text{ex},n_2}/2 +\text{h.t.},\\
\phi_{n_1}-\phi_{n_1-1} &= \phi_{n_1}' - \phi_{n_1}''/2 +\text{h.t.}= \mu_{\text{ex},n_1} - \frac{1}{2\beta} \frac{\Gamma'_{n_1}}{\Gamma_{n_1}} - \mu'_{\text{ex},n_1}/2+\text{h.t.}
\end{align}


\begin{thebibliography}{26}%
\makeatletter
\providecommand \@ifxundefined [1]{%
 \@ifx{#1\undefined}
}%
\providecommand \@ifnum [1]{%
 \ifnum #1\expandafter \@firstoftwo
 \else \expandafter \@secondoftwo
 \fi
}%
\providecommand \@ifx [1]{%
 \ifx #1\expandafter \@firstoftwo
 \else \expandafter \@secondoftwo
 \fi
}%
\providecommand \natexlab [1]{#1}%
\providecommand \enquote  [1]{``#1''}%
\providecommand \bibnamefont  [1]{#1}%
\providecommand \bibfnamefont [1]{#1}%
\providecommand \citenamefont [1]{#1}%
\providecommand \href@noop [0]{\@secondoftwo}%
\providecommand \href [0]{\begingroup \@sanitize@url \@href}%
\providecommand \@href[1]{\@@startlink{#1}\@@href}%
\providecommand \@@href[1]{\endgroup#1\@@endlink}%
\providecommand \@sanitize@url [0]{\catcode `\\12\catcode `\$12\catcode
  `\&12\catcode `\#12\catcode `\^12\catcode `\_12\catcode `\%12\relax}%
\providecommand \@@startlink[1]{}%
\providecommand \@@endlink[0]{}%
\providecommand \url  [0]{\begingroup\@sanitize@url \@url }%
\providecommand \@url [1]{\endgroup\@href {#1}{\urlprefix }}%
\providecommand \urlprefix  [0]{URL }%
\providecommand \Eprint [0]{\href }%
\providecommand \doibase [0]{http://dx.doi.org/}%
\providecommand \selectlanguage [0]{\@gobble}%
\providecommand \bibinfo  [0]{\@secondoftwo}%
\providecommand \bibfield  [0]{\@secondoftwo}%
\providecommand \translation [1]{[#1]}%
\providecommand \BibitemOpen [0]{}%
\providecommand \bibitemStop [0]{}%
\providecommand \bibitemNoStop [0]{.\EOS\space}%
\providecommand \EOS [0]{\spacefactor3000\relax}%
\providecommand \BibitemShut  [1]{\csname bibitem#1\endcsname}%
\let\auto@bib@innerbib\@empty
\bibitem [{\citenamefont {Ala-Nissila}\ \emph {et~al.}(2002)\citenamefont
  {Ala-Nissila}, \citenamefont {Ferrando},\ and\ \citenamefont {Ying}}]{ala}%
  \BibitemOpen
  \bibfield  {author} {\bibinfo {author} {\bibfnamefont {T.}~\bibnamefont
  {Ala-Nissila}}, \bibinfo {author} {\bibfnamefont {R.}~\bibnamefont
  {Ferrando}}, \ and\ \bibinfo {author} {\bibfnamefont {S.~C.}\ \bibnamefont
  {Ying}},\ }\bibfield  {title} {\enquote {\bibinfo {title} {Collective and
  single particle diffusion on surfaces},}\ }\href@noop {} {\bibfield
  {journal} {\bibinfo  {journal} {Advances in Physics}\ }\textbf {\bibinfo
  {volume} {51}},\ \bibinfo {pages} {949} (\bibinfo {year} {2002})}\BibitemShut
  {NoStop}%
\bibitem [{\citenamefont {Gomer}(1990)}]{gomer}%
  \BibitemOpen
  \bibfield  {author} {\bibinfo {author} {\bibfnamefont {R.}~\bibnamefont
  {Gomer}},\ }\bibfield  {title} {\enquote {\bibinfo {title} {Diffusion of
  adsorbates on metal surfaces},}\ }\href@noop {} {\bibfield  {journal}
  {\bibinfo  {journal} {Rep. Prog. Phys.}\ }\textbf {\bibinfo {volume} {53}},\
  \bibinfo {pages} {917--1002} (\bibinfo {year} {1990})}\BibitemShut {NoStop}%
\bibitem [{\citenamefont {Antczak}\ and\ \citenamefont
  {Ehrlich}(2010)}]{antczak}%
  \BibitemOpen
  \bibfield  {author} {\bibinfo {author} {\bibfnamefont {G.}~\bibnamefont
  {Antczak}}\ and\ \bibinfo {author} {\bibfnamefont {G.}~\bibnamefont
  {Ehrlich}},\ }\href@noop {} {\emph {\bibinfo {title} {Surface Diffusion,
  Metals, Metal Atoms, and Clusters}}}\ (\bibinfo  {publisher} {Cambridge
  University Press},\ \bibinfo {address} {Cambridge},\ \bibinfo {year}
  {2010})\BibitemShut {NoStop}%
\bibitem [{\citenamefont {Paul}\ \emph {et~al.}(2014)\citenamefont {Paul},
  \citenamefont {Laurila}, \citenamefont {Vuorinen},\ and\ \citenamefont
  {Divinski}}]{paul}%
  \BibitemOpen
  \bibfield  {author} {\bibinfo {author} {\bibfnamefont {A.}~\bibnamefont
  {Paul}}, \bibinfo {author} {\bibfnamefont {T.}~\bibnamefont {Laurila}},
  \bibinfo {author} {\bibfnamefont {V.}~\bibnamefont {Vuorinen}}, \ and\
  \bibinfo {author} {\bibfnamefont {S.~V.}\ \bibnamefont {Divinski}},\
  }\href@noop {} {\emph {\bibinfo {title} {Thermodynamics, Diffusion and the
  Kirkendall Effect in Solids}}}\ (\bibinfo  {publisher} {Springer},\ \bibinfo
  {address} {Heidelberg},\ \bibinfo {year} {2014})\BibitemShut {NoStop}%
\bibitem [{\citenamefont {Mehrer}(2007)}]{mehrer}%
  \BibitemOpen
  \bibfield  {author} {\bibinfo {author} {\bibfnamefont {H.}~\bibnamefont
  {Mehrer}},\ }\href@noop {} {\emph {\bibinfo {title} {Diffusion in Solids}}}\
  (\bibinfo  {publisher} {Springer},\ \bibinfo {address} {Berlin},\ \bibinfo
  {year} {2007})\BibitemShut {NoStop}%
\bibitem [{\citenamefont {Kutner}(1981)}]{kutner}%
  \BibitemOpen
  \bibfield  {author} {\bibinfo {author} {\bibfnamefont {R.}~\bibnamefont
  {Kutner}},\ }\bibfield  {title} {\enquote {\bibinfo {title} {Chemical
  diffusion in the lattice gas of non-interacting particles},}\ }\href@noop {}
  {\bibfield  {journal} {\bibinfo  {journal} {Physics Letters A}\ }\textbf
  {\bibinfo {volume} {81}},\ \bibinfo {pages} {239} (\bibinfo {year}
  {1981})}\BibitemShut {NoStop}%
\bibitem [{\citenamefont {Kehr}\ \emph {et~al.}(1981)\citenamefont {Kehr},
  \citenamefont {Kutner},\ and\ \citenamefont {Binder}}]{kehr}%
  \BibitemOpen
  \bibfield  {author} {\bibinfo {author} {\bibfnamefont {K.~W.}\ \bibnamefont
  {Kehr}}, \bibinfo {author} {\bibfnamefont {R.}~\bibnamefont {Kutner}}, \ and\
  \bibinfo {author} {\bibfnamefont {K.}~\bibnamefont {Binder}},\ }\bibfield
  {title} {\enquote {\bibinfo {title} {Diffusion in concentrated lattice gases.
  self-diffusion of noninteracting particles in three-dimensional lattices},}\
  }\href@noop {} {\bibfield  {journal} {\bibinfo  {journal} {Phys. Rev. B}\
  }\textbf {\bibinfo {volume} {23}},\ \bibinfo {pages} {4931} (\bibinfo {year}
  {1981})}\BibitemShut {NoStop}%
\bibitem [{\citenamefont {Hill}(1986)}]{hill}%
  \BibitemOpen
  \bibfield  {author} {\bibinfo {author} {\bibfnamefont {T.~L.}\ \bibnamefont
  {Hill}},\ }\href@noop {} {\emph {\bibinfo {title} {An Introduction to
  Statistical Thermodynamics}}}\ (\bibinfo  {publisher} {Dover},\ \bibinfo
  {address} {New York},\ \bibinfo {year} {1986})\BibitemShut {NoStop}%
\bibitem [{\citenamefont {Ferrando}\ and\ \citenamefont
  {Scalas}(1993)}]{ferrando}%
  \BibitemOpen
  \bibfield  {author} {\bibinfo {author} {\bibfnamefont {R.}~\bibnamefont
  {Ferrando}}\ and\ \bibinfo {author} {\bibfnamefont {E.}~\bibnamefont
  {Scalas}},\ }\bibfield  {title} {\enquote {\bibinfo {title} {Self-diffusion
  in a {2D} lattice gas with lateral interactions},}\ }\href@noop {} {\bibfield
   {journal} {\bibinfo  {journal} {Surface Science}\ }\textbf {\bibinfo
  {volume} {281}},\ \bibinfo {pages} {178} (\bibinfo {year}
  {1993})}\BibitemShut {NoStop}%
\bibitem [{\citenamefont {Hjelt}\ \emph {et~al.}(1997)\citenamefont {Hjelt},
  \citenamefont {Vattulainen}, \citenamefont {Merikoski}, \citenamefont
  {Ala-Nissila},\ and\ \citenamefont {Ying}}]{hjelt}%
  \BibitemOpen
  \bibfield  {author} {\bibinfo {author} {\bibfnamefont {T.}~\bibnamefont
  {Hjelt}}, \bibinfo {author} {\bibfnamefont {I.}~\bibnamefont {Vattulainen}},
  \bibinfo {author} {\bibfnamefont {J.}~\bibnamefont {Merikoski}}, \bibinfo
  {author} {\bibfnamefont {T.}~\bibnamefont {Ala-Nissila}}, \ and\ \bibinfo
  {author} {\bibfnamefont {S.~C.}\ \bibnamefont {Ying}},\ }\bibfield  {title}
  {\enquote {\bibinfo {title} {A dynamical mean field theory for the study of
  surface diffusion constants},}\ }\href@noop {} {\bibfield  {journal}
  {\bibinfo  {journal} {Surface Science}\ }\textbf {\bibinfo {volume} {380}},\
  \bibinfo {pages} {L501} (\bibinfo {year} {1997})}\BibitemShut {NoStop}%
\bibitem [{\citenamefont {Darken}(1948)}]{darken}%
  \BibitemOpen
  \bibfield  {author} {\bibinfo {author} {\bibfnamefont {L.~S.}\ \bibnamefont
  {Darken}},\ }\bibfield  {title} {\enquote {\bibinfo {title} {Diffusion,
  mobility and their interrelation through free energy in binary metallic
  systems},}\ }\href@noop {} {\bibfield  {journal} {\bibinfo  {journal} {Trans.
  AIME}\ }\textbf {\bibinfo {volume} {175}},\ \bibinfo {pages} {184} (\bibinfo
  {year} {1948})}\BibitemShut {NoStop}%
\bibitem [{\citenamefont {Widom}(1963)}]{widom}%
  \BibitemOpen
  \bibfield  {author} {\bibinfo {author} {\bibfnamefont {B.}~\bibnamefont
  {Widom}},\ }\bibfield  {title} {\enquote {\bibinfo {title} {Some topics in
  the theory of fluids},}\ }\href@noop {} {\bibfield  {journal} {\bibinfo
  {journal} {J. Chem. Phys.}\ }\textbf {\bibinfo {volume} {39}},\ \bibinfo
  {pages} {2808} (\bibinfo {year} {1963})}\BibitemShut {NoStop}%
\bibitem [{\citenamefont {Mart\'{\i}nez}\ and\ \citenamefont
  {Hoyuelos}(2018)}]{mdp}%
  \BibitemOpen
  \bibfield  {author} {\bibinfo {author} {\bibfnamefont {M.~Di~Pietro}\
  \bibnamefont {Mart\'{\i}nez}}\ and\ \bibinfo {author} {\bibfnamefont
  {M.}~\bibnamefont {Hoyuelos}},\ }\bibfield  {title} {\enquote {\bibinfo
  {title} {Mean-field approach to diffusion with interaction: Darken equation
  and numerical validation},}\ }\href@noop {} {\bibfield  {journal} {\bibinfo
  {journal} {Phys. Rev. E}\ }\textbf {\bibinfo {volume} {98}},\ \bibinfo
  {pages} {022121} (\bibinfo {year} {2018})}\BibitemShut {NoStop}%
\bibitem [{\citenamefont {Mart\'{\i}nez}\ and\ \citenamefont
  {Hoyuelos}(2019{\natexlab{a}})}]{dipietro2}%
  \BibitemOpen
  \bibfield  {author} {\bibinfo {author} {\bibfnamefont {M.~Di~Pietro}\
  \bibnamefont {Mart\'{\i}nez}}\ and\ \bibinfo {author} {\bibfnamefont
  {M.}~\bibnamefont {Hoyuelos}},\ }\bibfield  {title} {\enquote {\bibinfo
  {title} {From diffusion experiments to mean-field theory simulations and
  back},}\ }\href@noop {} {\bibfield  {journal} {\bibinfo  {journal} {J. Stat.
  Mech.: Theory Exp.}\ }\textbf {\bibinfo {volume} {2019}},\ \bibinfo {pages}
  {113201} (\bibinfo {year} {2019}{\natexlab{a}})}\BibitemShut {NoStop}%
\bibitem [{\citenamefont {Mart\'{\i}nez}\ and\ \citenamefont
  {Hoyuelos}(2019{\natexlab{b}})}]{dipietro3}%
  \BibitemOpen
  \bibfield  {author} {\bibinfo {author} {\bibfnamefont {M.~Di~Pietro}\
  \bibnamefont {Mart\'{\i}nez}}\ and\ \bibinfo {author} {\bibfnamefont
  {M.}~\bibnamefont {Hoyuelos}},\ }\bibfield  {title} {\enquote {\bibinfo
  {title} {Diffusion in binary mixtures: An analysis of the dependence on the
  thermodynamic factor},}\ }\href@noop {} {\bibfield  {journal} {\bibinfo
  {journal} {Phys. Rev. E}\ }\textbf {\bibinfo {volume} {100}},\ \bibinfo
  {pages} {022112} (\bibinfo {year} {2019}{\natexlab{b}})}\BibitemShut
  {NoStop}%
\bibitem [{\citenamefont {Kardar}(2007)}]{kardar}%
  \BibitemOpen
  \bibfield  {author} {\bibinfo {author} {\bibfnamefont {M.}~\bibnamefont
  {Kardar}},\ }\href@noop {} {\emph {\bibinfo {title} {Statistical Physics of
  Particles}}}\ (\bibinfo  {publisher} {Cambridge University Press},\ \bibinfo
  {address} {Cambridge},\ \bibinfo {year} {2007})\BibitemShut {NoStop}%
\bibitem [{\citenamefont {Hansen}\ and\ \citenamefont
  {McDonald}(2013)}]{hansen2}%
  \BibitemOpen
  \bibfield  {author} {\bibinfo {author} {\bibfnamefont {J.-P.}\ \bibnamefont
  {Hansen}}\ and\ \bibinfo {author} {\bibfnamefont {I.~R.}\ \bibnamefont
  {McDonald}},\ }\href@noop {} {\emph {\bibinfo {title} {Theory of Simple
  Liquids: With Applications to Soft Matter}}}\ (\bibinfo  {publisher}
  {Academic},\ \bibinfo {address} {Oxford},\ \bibinfo {year}
  {2013})\BibitemShut {NoStop}%
\bibitem [{\citenamefont {Gardiner}(1997)}]{gardiner}%
  \BibitemOpen
  \bibfield  {author} {\bibinfo {author} {\bibfnamefont {C.~W.}\ \bibnamefont
  {Gardiner}},\ }\href@noop {} {\emph {\bibinfo {title} {Handbook of Stochastic
  Methods}}},\ \bibinfo {edition} {2nd}\ ed.\ (\bibinfo  {publisher}
  {Springer},\ \bibinfo {address} {Berlin},\ \bibinfo {year}
  {1997})\BibitemShut {NoStop}%
\bibitem [{\citenamefont {M.~Tagliazucchi}(2017)}]{taglia}%
  \BibitemOpen
  \bibfield  {author} {\bibinfo {author} {\bibfnamefont {I.~Szleifer~(eds.}\
  \bibnamefont {M.~Tagliazucchi}},\ }\href@noop {} {\emph {\bibinfo {title}
  {Chemically Modified Nanopores and Nanochannels}}}\ (\bibinfo  {publisher}
  {Elsevier},\ \bibinfo {address} {Amsterdam},\ \bibinfo {year}
  {2017})\BibitemShut {NoStop}%
\bibitem [{\citenamefont {Su\'arez}\ \emph {et~al.}(2015)\citenamefont
  {Su\'arez}, \citenamefont {Hoyuelos},\ and\ \citenamefont
  {M\'artin}}]{suarez}%
  \BibitemOpen
  \bibfield  {author} {\bibinfo {author} {\bibfnamefont {G.}~\bibnamefont
  {Su\'arez}}, \bibinfo {author} {\bibfnamefont {M.}~\bibnamefont {Hoyuelos}},
  \ and\ \bibinfo {author} {\bibfnamefont {H.}~\bibnamefont {M\'artin}},\
  }\bibfield  {title} {\enquote {\bibinfo {title} {Mean-field approach for
  diffusion of interacting particles},}\ }\href {\doibase
  10.1103/PhysRevE.92.062118} {\bibfield  {journal} {\bibinfo  {journal} {Phys.
  Rev. E}\ }\textbf {\bibinfo {volume} {92}},\ \bibinfo {pages} {062118}
  (\bibinfo {year} {2015})}\BibitemShut {NoStop}%
\bibitem [{\citenamefont {Mart\'{\i}nez}\ \emph {et~al.}(2020)\citenamefont
  {Mart\'{\i}nez}, \citenamefont {Giuliano},\ and\ \citenamefont
  {Hoyuelos}}]{dipietro4}%
  \BibitemOpen
  \bibfield  {author} {\bibinfo {author} {\bibfnamefont {M.~Di~Pietro}\
  \bibnamefont {Mart\'{\i}nez}}, \bibinfo {author} {\bibfnamefont
  {M.}~\bibnamefont {Giuliano}}, \ and\ \bibinfo {author} {\bibfnamefont
  {M.}~\bibnamefont {Hoyuelos}},\ }\bibfield  {title} {\enquote {\bibinfo
  {title} {Out-of-equilibrium monte carlo simulations of a classical gas with
  bose-einstein statistics},}\ }\href@noop {} {\bibfield  {journal} {\bibinfo
  {journal} {Phys. Rev. E}\ }\textbf {\bibinfo {volume} {102}},\ \bibinfo
  {pages} {062125} (\bibinfo {year} {2020})}\BibitemShut {NoStop}%
\bibitem [{\citenamefont {Reed}\ and\ \citenamefont {Ehrlich}(1981)}]{reed}%
  \BibitemOpen
  \bibfield  {author} {\bibinfo {author} {\bibfnamefont {D.~A.}\ \bibnamefont
  {Reed}}\ and\ \bibinfo {author} {\bibfnamefont {G.}~\bibnamefont {Ehrlich}},\
  }\bibfield  {title} {\enquote {\bibinfo {title} {Surface diffusion, atomic
  jump rates and thermodynamics},}\ }\href@noop {} {\bibfield  {journal}
  {\bibinfo  {journal} {Surface Science}\ }\textbf {\bibinfo {volume} {102}},\
  \bibinfo {pages} {588} (\bibinfo {year} {1981})}\BibitemShut {NoStop}%
\bibitem [{\citenamefont {Gortel}\ and\ \citenamefont
  {Załuska-Kotur}(2004)}]{gortel}%
  \BibitemOpen
  \bibfield  {author} {\bibinfo {author} {\bibfnamefont {Z.~W.}\ \bibnamefont
  {Gortel}}\ and\ \bibinfo {author} {\bibfnamefont {M.~A.}\ \bibnamefont
  {Załuska-Kotur}},\ }\bibfield  {title} {\enquote {\bibinfo {title} {Chemical
  diffusion in an interacting lattice gas: Analytic theory and simple
  applications},}\ }\href@noop {} {\bibfield  {journal} {\bibinfo  {journal}
  {Phys. Rev. B}\ }\textbf {\bibinfo {volume} {70}},\ \bibinfo {pages} {125431}
  (\bibinfo {year} {2004})}\BibitemShut {NoStop}%
\bibitem [{\citenamefont {Badowski1}\ \emph {et~al.}(2010)\citenamefont
  {Badowski1}, \citenamefont {Załuska-Kotur},\ and\ \citenamefont
  {Gortel}}]{badowski}%
  \BibitemOpen
  \bibfield  {author} {\bibinfo {author} {\bibfnamefont {L}~\bibnamefont
  {Badowski1}}, \bibinfo {author} {\bibfnamefont {M.~A.}\ \bibnamefont
  {Załuska-Kotur}}, \ and\ \bibinfo {author} {\bibfnamefont {Z.~W.}\
  \bibnamefont {Gortel}},\ }\bibfield  {title} {\enquote {\bibinfo {title}
  {Collective diffusion in a non-homogeneous interacting lattice gas},}\
  }\href@noop {} {\bibfield  {journal} {\bibinfo  {journal} {J. Stat. Mech.:
  Theory Exp.}\ }\textbf {\bibinfo {volume} {2010}},\ \bibinfo {pages} {P03008}
  (\bibinfo {year} {2010})}\BibitemShut {NoStop}%
\bibitem [{\citenamefont {Mińkowski}\ and\ \citenamefont
  {Załuska-Kotur}(2015)}]{minkowski2}%
  \BibitemOpen
  \bibfield  {author} {\bibinfo {author} {\bibfnamefont {M.}~\bibnamefont
  {Mińkowski}}\ and\ \bibinfo {author} {\bibfnamefont {M.~A.}\ \bibnamefont
  {Załuska-Kotur}},\ }\bibfield  {title} {\enquote {\bibinfo {title}
  {Diffusion of {G}a adatoms at the surface of {G}a{A}s(001) $c(4 \times
  4)\alpha$ and $\beta$ reconstructions},}\ }\href@noop {} {\bibfield
  {journal} {\bibinfo  {journal} {Phys. Rev. B}\ }\textbf {\bibinfo {volume}
  {91}},\ \bibinfo {pages} {075411} (\bibinfo {year} {2015})}\BibitemShut
  {NoStop}%
\bibitem [{\citenamefont {Mińkowski}\ and\ \citenamefont
  {Załuska-Kotur}(2018)}]{minkowski}%
  \BibitemOpen
  \bibfield  {author} {\bibinfo {author} {\bibfnamefont {M.}~\bibnamefont
  {Mińkowski}}\ and\ \bibinfo {author} {\bibfnamefont {M.~A.}\ \bibnamefont
  {Załuska-Kotur}},\ }\bibfield  {title} {\enquote {\bibinfo {title}
  {Collective diffusion of dense adsorbate at surfaces of arbitrary
  geometry},}\ }\href@noop {} {\bibfield  {journal} {\bibinfo  {journal} {J.
  Stat. Mech.: Theory Exp.}\ }\textbf {\bibinfo {volume} {2018}},\ \bibinfo
  {pages} {053208} (\bibinfo {year} {2018})}\BibitemShut {NoStop}%
\end{thebibliography}
%

\end{document}